**Electro-Fluidic Shuttle Memory Device: Classical Molecular Dynamics Study**

Jeong Won K$_{\text{ANG}}$* and Ho Jung H$_{\text{WANG}}$
*Computational Semiconductor Laboratory, Department of Electronic Engineering, Chung-Ang University, 221 HukSuk-Dong, DongJak-Ku, Seoul 156-756, Korea*

Tel: 82-2-820-5296
Fax: 82-2-825-1584
E-mail: gardenriver@korea.com

We investigated the internal dynamics of several electro-fluid shuttle memory elements, consisting of several media encapsulated in $C_{640}$ nanocapsule. The systems proposed were

(i) bucky shuttle memory devices ($C_{36}^{+}$@$C_{420}$ and $C_{60}^{+}$@$C_{420}$),

(ii) encapsulated-ions shuttle memory devices ((13 $K^{+}$)@$C_{420}$, (3 $K^{+}$-$C_{60}$-2 $K^{+}$)@$C_{640}$ and (5 $K^{+}$-$C_{60}$)@$C_{640}$) and

(iii) *endo*-fullerenes shuttle memory devices (($K^{+}$@$C_{60}$-$F^{-}$@$C_{60}$)@$C_{640}$).

Energetics and operating responses of several electro-fluidic shuttle memory devices, such as transitions between the two states of the $C_{640}$ capsule, were examined by classical molecular dynamics simulations of the shuttle media in the $C_{640}$ capsule under the external force fields. The operating force fields for the stable operations of the shuttle memory device were investigated.

## 1. Introduction

Fullerene-related materials have attracted considerable attention this last decade due to their unique physical and chemical properties [1,2]. Compared to other carbon structures, fullerenes have revealed promising applications in a wide variety of very important technological processes such as in designing electronic devices, super-fibers, catalytic materials, etc [3]. Especially the large empty space (particularly inside carbon nanotubes (CNTs)) has open also new applications as storage materials with high capacity and stability [4]. These cavities are large enough to accommodate a wide variety of atomic and molecular species, the presence of which can significantly influence the properties of the materials. In particular, a new type of self-assembled hybrid structures called "nanopeapods", consisting of fullerene arrays inside single-walled CNTs, have recently been reported [5-10]. The application of nanopeapods ranges from nanometer-sized containers of chemical reactant [8] to date storage [11] and high temperature superconductor [12]. The encapsulation of fullerenes (such as $C_{60}$) in nanotubes is favorable on energetic grounds and occurs rapidly by exposing nanotubes to sublimed fullerenes. Kwon *et al* [11] reported that multi-walled nanotubes called "bucky shuttle" [13] were synthesized from elemental carbon under specific conditions, and investigated bucky shuttle memory device, which acted as nanometer-sized memory element, using molecular dynamics (MD) simulations.

While aligned bucky shuttle structures are difficult to be in self-assembly, nanopeapods can be synthesized in the aligned structures using bundles of single-walled CNTs. If some processes, such as $C_{60}$ intercalation control, nanolithography, carbon nanotube cutting and carbon nanotube capping, are treated appropriately to the aligned nanopeapods, the aligned bucky shuttles can be synthesized. Cioslowski and Fleischmann [14] were investigated the endohedral fullerene (*endo*-fullerene) complexes including $F^{-}$, Ne, $Na^{+}$, $Mg^{2+}$ and $Al^{3+}$. The $F^{-}$ complex was stabilized more than the $Na^{+}$ one, despite equal absolute values of the formal charges of the $F^{-}$ and $Na^{+}$ ions. A lot of *endo*-fullerenes and *endo*-fullerene-encapsulated carbon nanotubes have been synthesized and investigated by several experimental and theoretical methods, and these have been reviewed in Refs. 15-17. Potassium-intercalated carbon onions including potassium@fullerenes were synthesized and investigated in experiment [18]. Farajian *et al* [19] have reported the insertion of Na and K inside single-walled CNTs using *ab initio* calculations and MD simulations. Gao *et al* [20] calculated the optimum structure of intercalated K atoms in single-walled CNT crystals using classical MD and energy minimization. Miyamoto *et al* [21] also predicted the structures and the properties of a linear potassium-atom chain doped into a CNT by a density functional method. Liu Hong *et al* [22] investigated the energy barriers and the chirality effects of the in-tube doping of alkali metals in single-walled CNTs using classical MD based on the Lennard-Jones potentials. Jeong *et al* [24,25] attempted both experimental and theoretical investigations of Cs encapsulated in single-walled CNTs. The Cs intercalation was realized by ion irradiation to single-walled CNTs, and field-emission-type transmission electron microscopy and scanning transmission electron microscopy observations confirmed the encapsulated Cs configurations. Micro-fluidic memory devices were demonstrated through the use of an aqueous viscoelastic polymer solution as a working fluid [23].

From above previous works, we expect that electro-fluidic memory devices can be realized by nanotechnology based on carbon nanosciences. Therefore, in this paper, we investigate the operations and the properties of several proposed electro-fluidic shuttle memory devices composed of fullerenes and potassium ions using classical MD simulations.

## 2. Methods

For C–C interactions, we used the Tersoff-Brenner potential function that has been widely applied to carbon systems [26-28]. The long-range interactions were characterized with the Lennard-Jones 12-6 (LJ12-6) potential. For K–K and K–C interactions, we employed the force fields method developed and used for studying the K-intercalated fullerenes by Goddard III and co-authors [20,29,30]. They assumed that the K is fully ionized to $K^+$ and that the charges on the single-walled CNT are distributed uniformly. This assumption is consistent with the assumptions used in the derivation of the van der Waals parameter for $K^+ - K^+$ and $C - K^+$, and then the force field method has given accurate results for lattice parameter and other properties for K-intercalated graphite compounds. Although there is still a debate in the literature about charge transfer in graphite intercalation compounds [31], we also assume that the charge of the K atom is fully transferred to the single-walled CNT. It is obvious that the Lennard-Jones models for the interactions between $K^+$ ions and between C atom and $K^+$ ion are very simple and are difficult to describe the complicated systems. However, the Lennard-Jones model has applied to the interactions of materials with very low binding energy or van der Waals interactions. Since the binding energy of $K^+$ is very low and the interaction between $K^+$ atom and C atom in CNTs is similar with a van der Waals interaction, the LJ12-6 potential is adopted in our calculations. The parameters for $K^+ - K^+$ were optimized to yield the experimental geometry of $KC_{24}$ and those for $C - K^+$ were obtained from the geometric means between C and $K^+$ [29]. In the same assumption, $C - F^-$ interactions were also modeled. Since the binding energy of $F^-$ is very low and the interaction between $F^-$ ion and C atom in CNTs is similar with a van der Waals interaction, the LJ12-6 potential can be adopted in our calculations. The sum of the van der Waals C and the ionic $F^-$ radii is 3.06 Å and the sum of the van der Waals C and the neutral F radii is 3.24 Å. Therefore, the equilibrium distances between C and $F^-$ was selected by the mean distance, 3.15 Å, and the potential well depth, 0.0011 eV, was obtained from Ref. 14. The interaction between fullerenes was calculated by the Girifalco potential [32]. The interaction between $K^+$ and $F^-$ was not considered in this work.

Therefore, the combined expression used to calculate the energy of the system under the external field is

$$E_b = E_{ext} + E_{Tersoff-Brenner} + E_{vdWC-C} + E_{LJC-K} + E_{LJK-K} + E_{LJC-F} + E_{Girifalco}, \qquad (1)$$

where $E_b$ is the total potential energy, $E_{ext}$ is the one-body potential energy by the external field, and $E_{Tersoff-Brenner}$ and $E_{vdWC-C}$ are the covalent bond energy obtained from the Tersoff-Brenner potential function and the van der Waals energy obtained from the LJ12-6 potential function, respectively. $E_{LJC-K}$, $E_{LJK-K}$ and $E_{LJC-F}$ are the potential energies obtained from the LJ12-6 potential functions for $C - K^+$, $K^+ - K^+$ and $C - F^-$, respectively. $E_{Girifalco}$ is the Girifalco potential between $C_{60}$ molecules. $E_{vdWC-C}$ is only non-zero after the Tersoff-Brenner potential goes to zero, and we use the cubic spline functions and the parameters by Mao et al [33] as follows:

$$, \qquad (2)$$

where $r_{ij}$ is the interatomic distance, $c_{n,k}$ are cubic spline coefficients, $\kappa_S$ is the cut-off distance of the Tersoff-Brenner potential function, $R_M$ is 3.2 Å, and $R_B$ is 10 Å. The parameters the LJ12-6 potential for $C - C$, $C - K^+$ and $K^+ - K^+$ are listed in Table 1.

We used both steepest descent (SD) and MD methods. The MD simulations used the same MD methods as were used in our previous works [34-38]. The MD code used the velocity Verlet algorithm, a Gunsteren-Berendsen thermostat to control temperature and neighbor lists to improve computing performance. MD time step was $5 \times 10^{-4}$ ps.

## 3. Results and Discussion
### 3.1 Fullerene Shuttle

The both ends of outer capsule were composed of the halves of the $C_{240}$ fullerene and were connected seamlessly to the cylindrical portion of the capsule, a 27 Å long segment of the (10, 10) carbon nanotube. Figure 1 shows the energetics of the $C_{36}@C_{640}$ system, consisting of a $C_{36}$ encapsulated in a $C_{640}$ capsule. The $C_{36}@C_{640}$ was relaxed using a simulated annealing simulation. Position of $C_{36}$ inside $C_{640}$ capsule was displaced by the increasing or the decreasing of 0.1 Å along the axis of $C_{640}$ capsule and at each position, the $C_{36}$ was optimized by the SD method. The optimum structures to hold the $C_{36}$ molecule at an inter-wall distance of 3.4 ~ 4.0 Å. Obviously, the van der Waals interaction stabilized the $C_{36}$ molecule at either end of the capsule, where the contact area is largest. This is reflected in the potential energy variation as shown in figure 1. The minimum potential energies are found near the both ends of the capsule and the activation energy barrier is 1.15 eV in the $C_{640}$ capsule.

We performed the classical MD simulations of the switching operations from 'bit 0' (the left side of the $C_{640}$ capsule) to 'bit 1' (the right side of the $C_{640}$ capsule) in the $C_{36}@C_{640}$ memory element. The $C_{36}$ was initially located at the 'bit 0'. At time $t = 0$, a constant external force was applied along the axis of the outer capsule (the z-axis). The temperature of the $C_{640}$ capsule was controlled by the Gunsteren-Berendsen thermostat at 10 K to minimize the thermal fluctuation effects of the $C_{640}$ capsule. The $F_{ext}$ on the atoms of the $C_{640}$ capsule was zero whereas the first deviation of the $F_{ext}$ on the atoms of the $C_{36}$ was a constant value. This assumption means that an external field only accelerates the encapsulated $C_{36}$. We observed the central position variations of the $C_{36}$ along the *z*-axis as a function of the MD time with the external force applied to the $C_{36}$ fullerene. For the initial structure obtained from the simulated annealing simulation, when the external force per atom ($F_{ext}$) was above 0.009 eV/Å, the $C_{36}$ fullerene inside the $C_{640}$ capsule could be accelerated. When $F_{ext}$ was below 0.008 eV/Å, the $C_{36}$ fullerene was never accelerated. To systemically study, five cases with different initial structures were investigated. In any case, when $F_{ext}$ was above 0.011 eV/Å, the $C_{36}$ encapsulated in the $C_{640}$ capsule were always accelerated.

Since the external field during a long time can make a kinetic energy of the $C_{36}$ to exceed a collapsing impact of the $C_{36}$ or the $C_{640}$, there is no need to apply a constant switching field during the entire bit flip process. Kwon *et al* [11] showed that a 0.5 ps pulse of a 0.1 ~ 0.5 V/Å field was found to suffice to detach the $C_{60}^-$ ion from its stable position and thus to change the memory state. In our MD simulations, a 1.2 ~ 2.0 ps pulse of a 0.01 ~ 0.015 eV/Å per atom (0.36 ~ 0.54 eV/Å per $C_{36}$) was found adequately. We tested the $C_{36}$ shuttle operations in the $C_{640}$ capsule as a function of the external force. In figure 2, the turn-on field ('bit 0' $\rightarrow$ 'bit 1') was 0.015 eV/Å per atom during 1.25 ps whereas the turn-off field ('bit 1' $\rightarrow$ 'bit 0') was 0.013 eV/Å per atom during 1.75 ps. The $C_{36}$ returned by the turn-off field and was captured in the cap of the $C_{640}$ capsule by the suitable turn-off field. The most suitable field was the 1.75 ps pulse of the 0.013 eV/Å per atom (0.47 eV/Å per $C_{36}$) in our MD simulations. The $C_{36}$ was not return by the weak turn-off field of 0.008 eV/Å per atom (0.324 eV/Å per $C_{36}$) in our MD simulations during 1.25 ps. Under a strong turn-off field of 0.016 eV/Å per atom (0.576 eV/Å per $C_{36}$) during 1.75 ps, the $C_{36}$ returned but three carbon atoms of the $C_{36}$ made the covalent bonds with three atoms of the $C_{640}$ capsule. Under a strong turn-off field of 0.016 eV/Å per atom (0.576 eV/Å per $C_{36}$) during 1.25 ps, the $C_{36}$ returned but it rebounded the other side of the $C_{640}$ capsule. Therefore, these cases could not be used to the shuttle memory operation. For the shuttle of the $C_{36}$ inside the $C_{640}$ capsule, the exact classification between 'bit 1' and 'bit 0' is necessary about 10 ps to stabilize. Therefore, the switching frequency is close to 0.1 THz and the data throughput rate is maximally about 100 Gbyte/s.

We also investigated a $C_{60}@C_{640}$ shuttle system as shown in figures 3 and 4. In figure 3, the $C_{60}$ initially located in the center of the $C_{640}$ capsule. The $F_{ext}$ on the atoms of the $C_{640}$ capsule was zero whereas the $F_{ext}$ on the atoms of the $C_{60}$ was a constant value. This assumption means that an external field only accelerates the encapsulated $C_{60}$. The central position variations of $C_{60}$ are shown as functions of MD time and external force. The turn-on field was 0.0025 eV/Å (0.15 eV/Å per $C_{60}$) per atom during 29.5 ps whereas the turn-off field was 0.0025 eV/Å per atom during 20 ps. Since the weak field was applied, a constant switching field was applied during the entire bit flip process. As mentioned above, there is no need to apply a constant switching field during the entire bit flip process. In figure 4, the $C_{60}$ initially located in the left cap of the $C_{640}$ capsule. The turn-on field was 0.015 eV/Å per atom (0.8 eV/Å per $C_{60}$) during 1.25 ps whereas the turn-off field was 0.012 eV/Å per atom (0.72 eV/Å per $C_{60}$) during 1.25 ps. Figure 4 shows the central position variation of the $C_{60}$, the total potential energy variation of the $C_{60}@C_{640}$ and the kinetic energy variation of the $C_{60}$ as functions of the external force and MD time. By turn-on field, the entire flip process was achieved in 10 ps, whereas by turn-off field, it was achieved in 25 ps. The kinetic energy of $C_{60}$ was increased under the external field and it was gradually decreased by the interaction between the $C_{60}$ molecule and the $C_{640}$ capsule wall as soon as the external field disappeared. However, when the $C_{60}$ was captured in the cap of the $C_{640}$ capsule, the attractive interaction by the large contact area of caps accelerated the $C_{60}$ as indicated arrows in figure 4(c). The total potential variation of system (figure 4(b)) is closely related to the kinetic energy variation of the $C_{60}$ (figure 4(c)). The minimum potential energies are found near the caps of the capsule and the activation energy barrier, 1.52 eV, can be obtained from the difference between the minimum potential energy in the cap and the maximum potential energy during the bit flip process. The response of $C_{60}$ under the applied external force fields is also similar to that of $C_{36}$ in the $C_{640}$ capsule.

### 3.2 Encapsulated-Ion Shuttles

Recently, the alkali ion encapsulations were realized by ion irradiation to single-walled CNTs, and field-emission-type transmission electron microscopy and scanning transmission electron microscopy observations confirmed the encapsulated configurations [24,25]. Therefore, the $C_{640}$ capsule encapsulating alkali ions was considered in a shuttle memory element. Figure 5 shows the energetics of the $(13K^+)@C_{640}$ system, consisting of thirteen $K^+$ ions encapsulated in the $C_{640}$ capsule. The $(13K^+)@C_{640}$ was relaxed using a simulated annealing simulation at the left side of the $C_{640}$. The optimized structure of $13K^+$ was an icosahedron as well known. Since the $C_{240}$ fullerene is an icohahedron, the encapsulated $13K^+$ are most stabilized as the icosahedral cluster such as $Na_{13}@C_{240}$ [39]. Positions of $13K^+$ inside $C_{640}$ capsule were displaced by 0.1 Å along the axis of the $C_{640}$ capsule and at each position, the $13K^+$ were optimized by the SD method. Obviously, the interactions between $K^+$ ion and C atom stabilized the $13K^+$ at either end of the capsule and the minimum potential energies are found near the ends of the capsule, where the contact area is largest, as shown in figure 3. The potential-well-depth of $13K^+$ is 0.121 eV in the $C_{640}$ capsule.

We performed the classical MD simulations of the switching operations from 'bit 0' to 'bit 1' in the $(13K^+)@C_{640}$ memory element. The $13K^+$ was initially located at the 'bit 0'. At time *t* = 0, a constant external force was applied along the axis of the outer capsule. The external force ($F_{ext}$) on the atoms of the $13K^+$ was a constant value. This assumption means that an external field only accelerates the encapsulated $13K^+$. We observed the mean position variations of the $13K^+$ along the *z*-axis as a function of the MD time with the external force applied to the $13 \text{ K}^+$. Since the binding energy between $13 \text{ K}^+$ and the cap of the outer capsule is very low, the turn-on field is also very low, ~0.001 eV/Å per ion. Therefore, there is also need to apply a constant switching field to the maintenance of the bit classification as well as during the entire bit flip process.

MD simulations were performed for the $13 \text{ K}^+$ shuttle operations in the $C_{640}$ capsule when the applied external forces were 0.01 and 0.014 eV/Å per ion. In figure 6, after the turn-on field ('bit 0' → 'bit 1'), when the external fields were not applied, the mean positions of $13 \text{ K}^+$ ions did not maintain the 'bit 1' position, i.e. $13 \text{ K}^+$ ions were not captured at the cap of the $C_{640}$ capsule in our MD simulations. This reason is that the binding energies between $\text{K}^+$ ions and the cap of the outer capsule are very low as mentioned above. The operation dynamics of the $(13 \text{ K}^+)@C_{640}$ shuttle memory device is very similar to those of the $C_{60}@C_{640}$ and $C_{36}@C_{640}$ shuttle memory device, except for the binding energies at the cap of the outer capsule. If the binding energy between an encapsulated ion and a carbon nanocapsule is as high as applied to the shuttle memory device, we expect that encapsulated-ion shuttle memory device can be in applications. Figure 7 shows the atomic structures in the processes of the $(13 \text{ K}^+)@C_{640}$ shuttle. Figures 7(a)-7(c) shows the turn-on process and figures 7(d)-7(f) shows the turn-off process.

Recent works have shown the intercalations of both ions and $C_{60}$ molecules into carbon nanotubes [24,25]. These results can make a proposal of different-type shuttle memory devices based on carbon nanocapsules. Figure 8 shows the response of a $(3 \text{ K}^+\text{-}C_{60}\text{-}2 \text{ K}^+)@C_{640}$ shuttle system as a function of applied external force. The external field accelerated five $\text{K}^+$ ions whereas external field did not affect the $C_{60}$ molecule. Turn-on and turn-off force fields were 0.002 eV/Å per atom during 29.5 and 20 ps, respectively. Under the changes of the external fields, both $\text{K}^+$ ions and $C_{60}$ alternatively shuttled 'bit 0' and 'bit 1'. Since three $\text{K}^+$ ions and two $\text{K}^+$ ions were initially divided by the $C_{60}$ molecule, the left side three $\text{K}^+$ ions did not meet the right side two $\text{K}^+$ ions. The central position variations of $C_{60}$ molecule were exactly distinguished in 'bit 0' and 'bit 1' whereas in the mean position variations of $\text{K}^+$ ions, it was difficult to divide bits. In our MD simulations with the low external fields, $\text{K}^+$ ions have never penetrated $C_{60}$ molecule.

Figure 9 shows the response of a $(5 \text{ K}^+\text{-}C_{60})@C_{640}$ shuttle system as a function of MD time with applied external force. The external field only accelerated five $\text{K}^+$ ions and the $C_{60}$ molecule was assumed by neutral molecule. The accelerated $\text{K}^+$ ions by the turn-on field pushed the $C_{60}$ molecule toward the right cap of the $C_{640}$ capsule. However, under the turn-off field, the $\text{K}^+$ ions were captured in the left cap of the $C_{640}$ capsule whereas the $C_{60}$ molecule remained in the right cap of the $C_{640}$ capsule because of the binding energy between the $C_{60}$ molecule and the cap of the $C_{640}$ capsule. When non-contacting or contacting between the $\text{K}^+$ ions and the $C_{60}$ molecule divides the 'bit 0' or the 'bit 1', this system as shown in figure 9 can operate a nanocapsule memory device.

### 3.3 Bipolar *Endo*-Fullerenes Shuttle

Cioslowski and Fleischmann [14] investigated the *endo*-fullerene complexes including $\text{F}^-$, Ne, $\text{Na}^+$, $\text{Mg}^{2+}$ and $\text{Al}^{3+}$. The $\text{F}^-$ complex was stabilized more than the $\text{Na}^+$ one, despite equal absolute values of the formal charges of the $\text{F}^-$ and $\text{Na}^+$ ions. The nanopeapods with *endo*-fullerenes have been widely investigated [15-17]. Therefore, bipolar endo-fullerenes shuttle memory device, including both anionic and cationic *endo*-fullerenes, was proposed and investigated. We performed the MD simulations for $(\text{K}^+@C_{60}\text{-}\text{F}^-@C_{60})@C_{640}$ shuttle system as shown in figures 10 and 11.

Figure 10 shows the central position variations of $C_{60}$ molecules as functions of MD time and applied external force. For A case, $\pm 1.5$ eV/Å pulses during 0.5 ps were applied. All carbon atoms were assumed by the neutral but the external field applied to only $\text{K}^+$ and $\text{F}^-$ ions. Position variations of $\text{K}^+$ and $\text{F}^-$ are shown along the central position variations of $C_{60}$. In all cases A-C, two *endo*-fullerenes were accelerated by the applied positive fields and collided with each other whereas they were not separated under a weak force field because of the binding energy between *endo*-fullerenes. When a proper external field is applied, the binding between *endo*-fullerenes can be broken. Figure 11 shows a case of the operation of $(\text{K}^+@C_{60}\text{-}\text{F}^-@C_{60})@C_{640}$ shuttle system, and $\text{K}^+@C_{60}$ and $\text{F}^-@C_{60}$ were initially located at both caps of the $C_{640}$ capsule. Two *endo*-fullerenes formed a dimer molecule at the right cap by the turn-on field. To stabilize two *endo*-fullerenes, a weak force field of 0.07 eV/Å during 10 ps was applied after the turn-on force field of 0.56 eV/Å during 4 ps. To break the bonding between two *endo*-fullerenes, a high turn-off force field of 1.75 eV/Å during 0.5 ps was applied. In the turn-off case, a weak force field of 0.07 eV/Å during 9.5 ps was applied after the turn-off force field to stabilize two *endo*-fullerenes at both caps. The responses of the shuttle operations can be confirmed with both the total potential energy variation and the inter-distance between two *endo*-fullerenes. In this system, the exact classification between 'bit 1' and 'bit 0' was necessary about 20 ps to stabilize.

### 4. Summary and Future Works

We studied the energetics and the operations of several electro-fluidic shuttle memory devices using classical molecular dynamics simulations. The proposed systems in this paper were the bucky shuttle memory devices ($C_{36}^+@C_{420}$ and $C_{60}^+@C_{420}$), the encapsulated-ion shuttle memory devices (($13 \text{ K}^+)@C_{420}$, $(3 \text{ K}^+\text{-}C_{60}\text{-}2 \text{ K}^+)@C_{640}$ and $(5 \text{ K}^+\text{-}C_{60})@C_{640}$) and the *endo*-fullerene shuttle memory device ($(\text{K}^+@C_{60}\text{-}\text{F}^-@C_{60})@C_{640}$). The switching processes from 'bit 0' to 'bit 1' in the electro-fluidic shuttle memory elements were investigated for several cases of the external force fields. The lowest energy configurations were found in the both caps of the $C_{640}$ capsule. Obviously, the interactions between capsule and shuttle media stabilized the shuttle media at caps of the capsule where the contact area was largest. Therefore, the bit flops could be classified with both the position of the shuttle media and the potential energy of the system. Classical molecular dynamics simulations showed that the electro-fluidic shuttle memory devices, including the bucky shuttle memory devices and encapsulated-ion shuttle memory device, could be operated by an adequate external force field.

This work shows that other nanocapsles based on the other nanotubes, such as boron nitride nanotubes and silicon oxide nanotubes, can be applied to the electro-mechanical shuttle memory devices. Since nanopeapods of boron nitride nanotubes were prepared [40], our future work will present the fullerene shuttle memory device based on an insulating boron nitride nanotube.

**TABLES**

Table 1. Parameters of Lennard-Jones 12-6 potential for $C-C$, $K^+ - K^+$ and $C-K^+$ atoms [26].

|  | $s \cdot 2^{1/6}$ (Å) | $e$ (kcal/mol) |
|---|---|---|
| $C-C$ | 3.8050 | 0.0692 |
| $K^+ - K^+$ | 4.0010 | 0.0700 |
| $C - K^+$ | 3.9018 | 0.0696 |

**FIGURE CAPTIONS**

Figure 1. Total potential energy of $C_{36}@C_{640}$ as a function of $C_{36}$ position.

Figure 2. Response of $C_{36}$ in the condition that turn-on and turn-off force fields were 0.015 and 0.013 eV/Å per atom during 1.25 and 1.75 ps, respectively.

Figure 3. Response of $C_{60}$ in the condition that turn-on and turn-off force fields were 0.0025 eV/Å per atom during 29.5 and 20 ps, respectively.

Figure 4. Response of $C_{60}$ in the condition that turn-on and turn-off force fields were 0.015 and 0.012 eV/Å per atom during 1.25 ps per atom, respectively. (a) Central position of $C_{60}$, (b) total potential energy of the system, (c) kinetic energy of $C_{60}$ and (d) applied external force as a function of MD time.

Figure 5. Total potential energy of $(13\,K^+)@C_{640}$ as a function of mean position of 13 $K^+$.

Figure 6. Responses of 13 $K^+$ in the condition that turn-on and turn-off force fields. Solid and dashed lines indicate the force fields of 0.015 and 0.01 eV/Å per ion, respectively.

Figure 7. Atomic structures in the processes of the $(13\,K^+)@C_{640}$ shuttle memory element. (a)-(c) shows the turn-on process and (d)-(f) shows the turn-off process.

Figure 8. Response of a $(3\,K^+ - C_{60} - 2\,K^+)@C_{640}$ shuttle system as a function of applied external force. Turn-on and turn-off force fields were 0 0.002 eV/Å per ion during 29.5 and 20 ps, respectively.

Figure 9. Response of a $(5\,K^+ - C_{60})@C_{640}$ shuttle system as a function of applied external force. Turn-on and turn-off force fields were 0.02 eV/Å per ion during 24.5 and 25 ps, respectively.

Figure 10. Response of a $(K^+@C_{60} - F^-@C_{60})@C_{640}$ shuttle system as a function of applied external force. Turn-on and turn-off force fields were 0.02 eV/Å per ion during 24.5 and 25 ps, respectively.

Figure 11. Response of a $(K^+@C_{60} - F^-@C_{60})@C_{640}$ shuttle system as a function of applied external force.

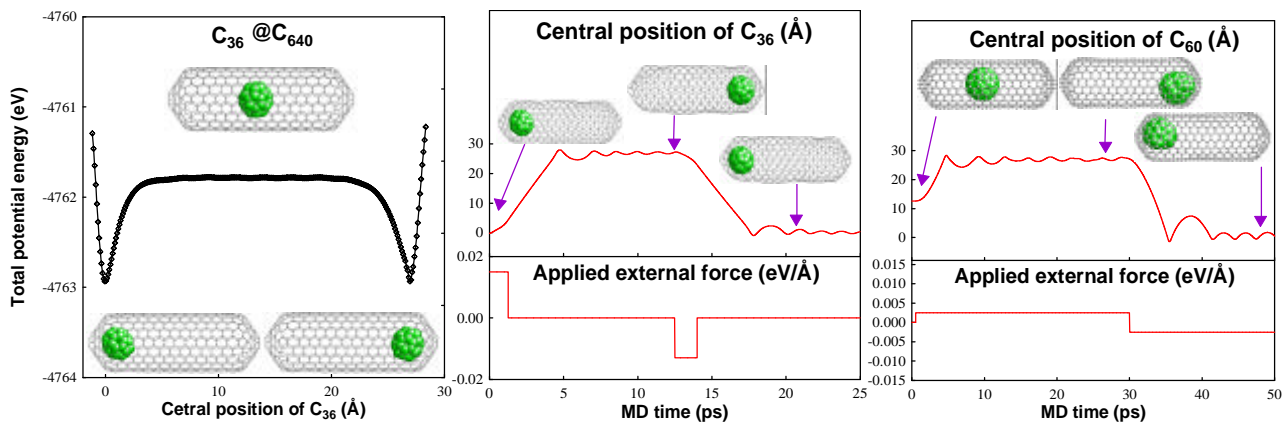

**FIGURE 1**     **FIGURE 2.**     **FIGURE 3.**

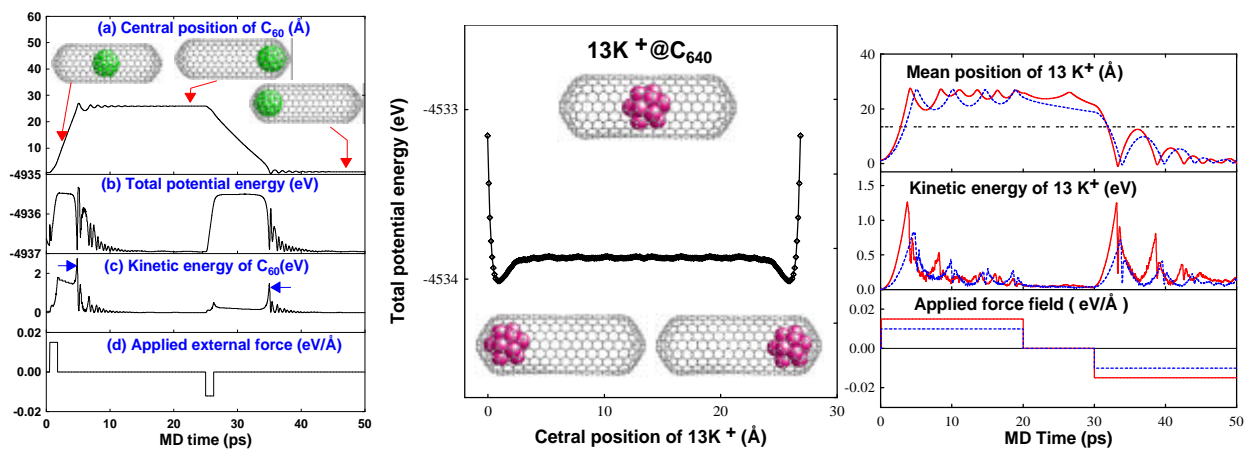

**FIGURE 4.**     **FIGURE 5.**     **FIGURE 6.**

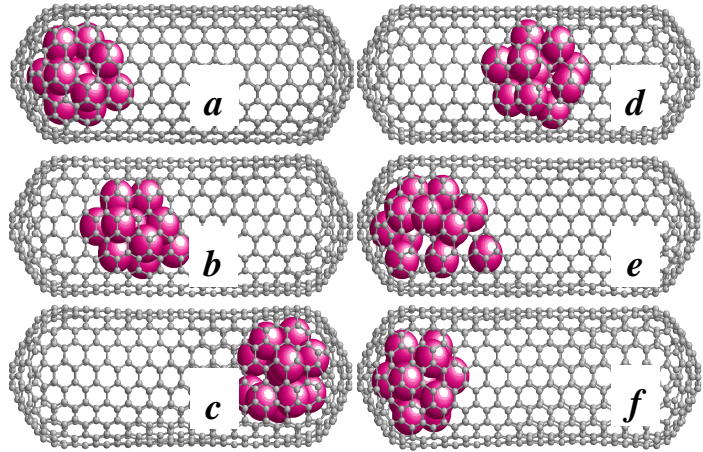

**FIGURE 7.**

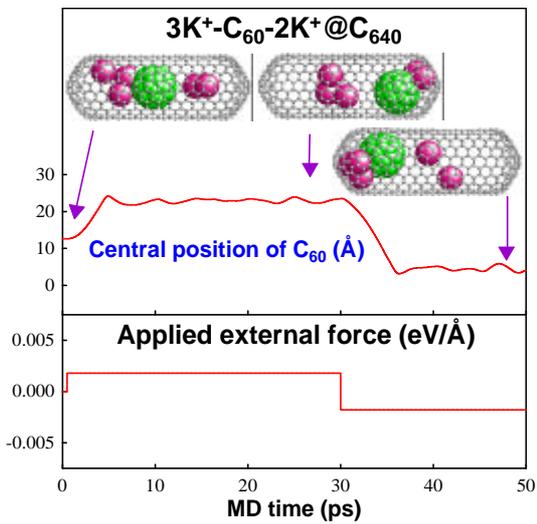

**FIGURE 8.**

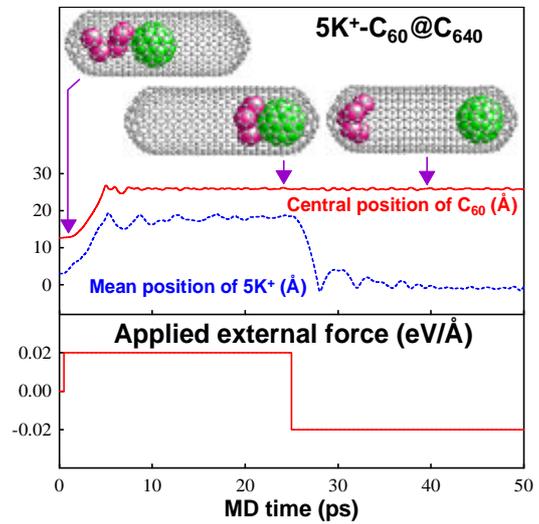

**FIGURE 9.**

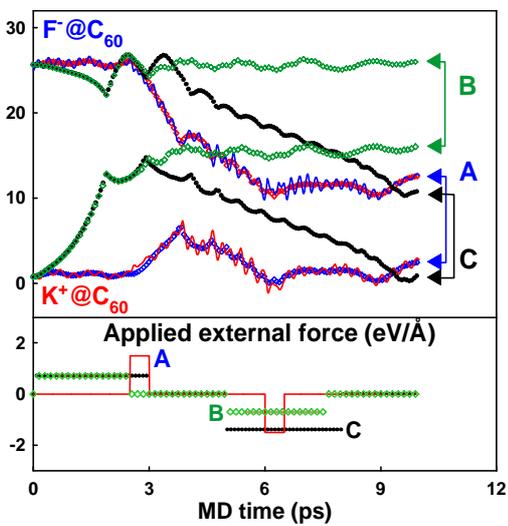

**FIGURE 10.**

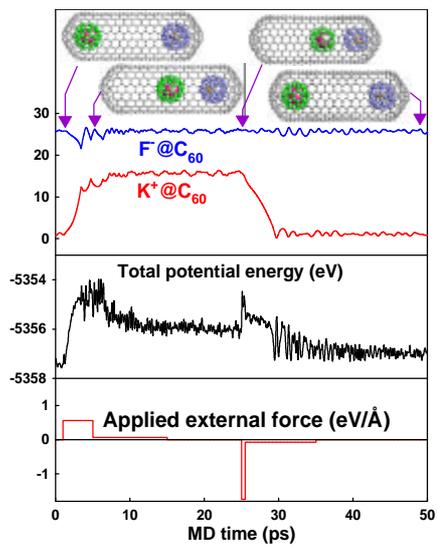

**FIGURE 11.**

7